# Memory Tagging and how it improves C/C++ memory safety

Kostya Serebryany, Evgenii Stepanov, Aleksey Shlyapnikov, Vlad Tsyrklevich, Dmitry Vyukov
Google
February 2018

# Introduction

Memory safety in C and C++ remains largely unresolved.

A technique usually called "memory tagging" may dramatically improve the situation if implemented in hardware with reasonable overhead. This paper describes two *existing* implementations of memory tagging: one is the full hardware implementation in SPARC; the other is a partially hardware-assisted compiler-based tool for AArch64. We describe the basic idea, evaluate the two implementations, and explain how they improve memory safety.

This paper is intended to initiate a wider discussion of memory tagging and to motivate the CPU and OS vendors to add support for it in the near future.

# Memory Safety in C/C++

C and C++ are well known for their performance and flexibility, but perhaps even more for their extreme memory unsafety. This year we are celebrating the 30th anniversary of the [Morris Worm](), one of the first known exploitations of a memory safety bug, and the problem is still not solved. If anything, it's more severe due to the exponential increase in the amount and complexity of modern software and a new focus on client-side applications with rich attack surfaces.

There are numerous tools (e.g. [AddressSanitizer]() or [Valgrind]()) and techniques (e.g. [fuzzing]() or [concolic execution]()) that find memory safety bugs during testing. There are also many techniques and tools that mitigate some aspects of memory safety bugs in production (e.g. [ASLR](), [CFI]()). Yet there is no end to memory safety bugs found in shipped software, and to exploits based on those bugs. The more constrained environments motivated the invention of new exploitation techniques (e.g. [ROP](), [DOP](), [JOP]()).

## AddressSanitizer

[AddressSanitizer]() (ASAN) is a software-only tool based on compiler instrumentation introduced to the [LLVM]() and [GCC]() compilers in 2011. ASAN finds the following classes of bugs:
- Temporal: [heap-use-after-free](), [stack-use-after-return](), [stack-use-after-scope]().
- Spatial: [heap-buffer-overflow](), [stack-buffer-overflow](), [global-buffer-overflow](), [container-overflow]().
- Other (not relevant for this discussion)

Among the bug classes not detected by ASAN are use-of-uninitialized-memory (see [MemorySanitizer]()) and [intra-object-buffer-overflow]().

*Redzones* around heap, stack, and global objects are used to detect spatial bugs, and thus an access that jumps over a redzone (aka non-linear buffer-overflow) may be undetected by ASAN.

*Quarantine* for heap and stack objects delays the reuse of deallocated objects and thus helps to detect temporal bugs. Exhausting the quarantine may lead to undetected temporal bugs.
A *shadow memory* maps every 8 bytes of the application memory into 1 byte of metadata to mark memory as either valid or invalid. The shadow is checked by instrumentation code injected by the compiler. The overheads are:
- 1.5x-3x in CPU, from instrumenting all loads and stores and from heavier malloc.
- 1.5x-3x in RAM, from redzones, quarantine, shadow, and internal bookkeeping.
- 1.5x-3x in Code Size, from instrumenting all loads and stores.

ASAN is heavily used at Google and across the industry and it has [found tens of thousands bugs](#). However, ASAN's usage outside of the development process (testing and fuzzing) is limited, mostly due to its combined CPU/RAM/Size overhead.

Other limitations of ASAN:
- Does not currently instrument assembly code
- Does not currently detect bugs in pre-compiled binaries
- Does not currently detect when the kernel accesses invalid user-space memory

ASAN is not a security hardening tool: the redzones and quarantine are easy to bypass.

# Memory Tagging

The general idea of **memory tagging** (MT, also known as memory coloring, memory tainting, lock and key) on 64-bit platforms is as follows:
- Every **TG** (tagging granularity) bytes of memory aligned by TG are associated with a tag of **TS** (tag size) bits. These TG bytes are called the **granule**.
- TS bits in the upper part of every pointer contain a tag.
- Memory allocation (e.g. malloc) chooses a tag, associates the memory chunk being allocated with this tag, and sets this tag in the returned pointer.
- Every load and store instruction raises an exception on mismatch between the pointer and memory tags.
- The memory access and tag check do not necessarily occur atomically with respect to each other.

The value of TS should be large enough to allow a sufficient number of different tag values (i.e. at least 4 to provide ~ 16 tag values) and small enough to fit into the unused upper bits of a pointer (usually up to 8 or 16, depending on the architecture). TS also affects the complexity and the size of the tag storage.

The value of TG is a balance between the size of the tag storage, the alignment requirement, and hardware possibilities.

Temporal and spatial bugs are detected probabilistically; if the tags are assigned to objects randomly, the probability of catching a bug is roughly $(2^{TS}-1)/2^{TS}$, i.e. even with TS=4 the probability is 15/16 or ~ 94%.

In the following sections we discuss two MT implementations:
- SPARC ADI (TG=64, TS=4)
- AArch64 HWASAN (TG=16, TS=8).

## SPARC ADI

The SPARC ADI hardware extension ([1], [2], [3]) is supported on SPARC M7/M8 CPUs running Solaris OS. There is also some indication that Linux support is in progress.
Main features of SPARC ADI:
- TG=64 and TS=4, i.e. every 64 bytes of memory are associated with a 4-bit tag.
- 2 (in SPARC M8) and 3 (in SPARC M7) tag values are reserved.
- The memory tag for a single 64-byte region is set with a single instruction:
  - `stxa  %r1, [ %r2 ] (144)`
- ADI supports setting a memory tag and zeroing the memory itself with one instruction:
  - `stxa  %r1, [ %r2 ] (146)  # %r1 must contain 0`
- Load/store generates SEGV on tag mismatch
- ADI has two modes with regard to handling store instructions: precise (slow, generates the exception exactly at the faulty instruction) and imprecise (faster, generates the exception at some later point). Load instructions are always handled precisely.
- Memory tags do not seem to be separately addressable, they are probably stored in extra ECC memory bits or in some other hidden hardware state. The only way to read the memory tag is via a special instruction.
- Untagged memory (with tag 0) can be accessed with tagged and untagged pointers.
- Memory has to be mapped with MAP_ADI to be tagged.
- Applying a memory tag to a non-resident (freshly mmap-ed) page makes this page resident.
- Syscalls return an error code if accessing a user provided buffer causes a tag mismatch.

## AArch64 HWASAN

HWASAN (hardware-assisted ASAN) is an AArch64-specific compiler-based tool.
- TG=16, the memory tags are stored in a directly mapped 16=>1 virtual shadow region allocated with mmap(MAP_NORESERVE), similar to ASAN.
- TS=8, the address tag is stored in the top address byte, leveraging the AArch64 hardware feature top-byte-ignore.
- Heap memory and pointers are tagged by a custom malloc implementation.
- Stack memory and pointers are tagged by compiler-injected code in the function prologue/epilogue.
- Checks for loads/stores are made by compiler-injected code.

HWASAN can be seen as an improved variant of ASAN available on AArch64; it also serves as a prototype for fully hardware-assisted memory tagging.

It is theoretically possible to implement HWASAN on architectures that do not have top-byte-ignore using page aliasing (mmap(MAP_SHARED)) to store the address tag in the meaningful address bits. However our experiments indicate that such implementation would be impractical due to huge stress on the TLB.

## Compiler And Run-time Support

Memory tagging hardware will not magically make C++ safer - it still requires cooperation between the compiler and the run-time.

Detecting **heap-related memory bugs** requires changes in malloc and free:
Malloc:
- Align the allocations by TG.
- Choose a tag T for every allocation.
- Tag the memory for the just-allocated chunk with T.
- Return the address tagged with T.

Free:
- Optionally retag the free-d memory with another tag.

Strategies for choosing the tag during malloc may vary. One such strategy is to use pseudo random tags. Another strategy would ensure that no two adjacent heap chunks have the same tag (for 100% reliable linear buffer overflow/underflow detection).

Detecting **stack-related memory bugs** requires changes in the compiler: the function prologue will need to align all local variables by TG, tag their locations with random tags, and the tagged pointers will need to be stored in separate (virtual) registers. Optionally, the function epilogue will retag the memory. Variations of this strategy are possible, e.g. to reduce register pressure or to enable stack-use-after-scope detection.

Detecting **buffer overflows in global variables** also requires compiler support (the details are out of scope of this paper).

# Overhead

## RAM

Major sources of **RAM overhead**:

- Over-aligning heap objects from the natural alignment (usually, 8 or 16) to TG
- Over-aligning stack objects
- Memory tag storage: TS extra bits per every TG bytes

We have measured overhead from heap over-alignment on a variety of 64-bit applications: one important server-side Google application, the Google Chrome browser on Linux, a set of 7 Android applications, and the Linux Kernel for arm64. The overheads are given in %, assuming the base is 8-byte alignment.

| Application / Heap alignment | 16 | 32 | 64 |
|---|---|---|---|
| A Google server-side application | 2% | 5.5% | 17% |
| Google Chrome for Linux | 0% | 7% | 28% |
| 7 Android Applications | 0%-6% | 3%-23% | 6%-42% |
| Linux kernel for arm64 | 4% | 4% | 14% |

We have also measured the average stack frame increase due to stack over-alignment on a subset of the above benchmarks:

| Application / Stack alignment | 16 | 32 | 64 |
|---|---|---|---|
| Google Chrome | 3.5% | 12% | 31% |
| Linux kernel for arm64 | 9% | 30% | 52% |

Even though larger tagging granularity requires less storage for the tags, it costs more due to heap and stack over-alignment. **Conclusion**: the optimal granularity is 16, assuming we need always-on MT and stack instrumentation. Only tagging the heap, especially with allocation-level sampling, will work fine with larger granularities.

## CPU

Major sources of **CPU overhead**:
- Tagging the heap/stack objects during allocation/deallocation.
- Increased working set size due to extra RAM consumption.
- Extra instructions to tag stack variables.
- Checking the tags during loads and stores - only for compiler-based instrumentation!

The CPU overhead of HWASAN is roughly 2x (similar to ASAN, which typically has 1.5x-3x CPU overhead [1]) and is mostly caused by extra instructions inserted at compile-time before loads and stores.

In contrast, with SPARC ADI the actual tag check during loads/stores is performed by hardware and introduces relatively low CPU overhead in the imprecise mode. We do not have a toolchain that instruments stack frames with SPARC ADI, so our measurements reflect heap-only mode.

We have measured the performance of SPARC ADI on the SPECint 2006 benchmarks by linking them against a simple ADI-enabled malloc, based on scudo malloc. A SPARC S7-2 server was used. We measured 6 configurations of malloc:
1. 16-byte alignment (base)
2. 64-byte alignment
3. 64-byte alignment + ADI tagging in malloc (equivalent to adi_set_version)
4. 64-byte alignment + ADI tagging+zeroing in malloc (equivalent to adi_memset)
5. 64-byte alignment + ADI tagging+zeroing in malloc + retagging in free
6. 64-byte alignment + ADI tagging+zeroing in malloc + precise trap mode for writes

Configurations 3 and 4 have shown the same results, i.e. zeroing out the memory has no additional cost. The following graph compares the 1$^{st}$ (base) configuration with configurations 2, 4, and 5. The gcc benchmark is the only benchmark where ADI tagging introduces noticeable overhead, which comes from a large number malloc calls with large sizes. Xalancbmk slows down by ~17% due to 64-byte over-alignment and it also gains ~20% in RSS; ADI tagging adds very little overhead on top of it. Two benchmarks, astar and omnetpp, are positively affected by 64-byte alignment and are mostly unaffected by ADI.

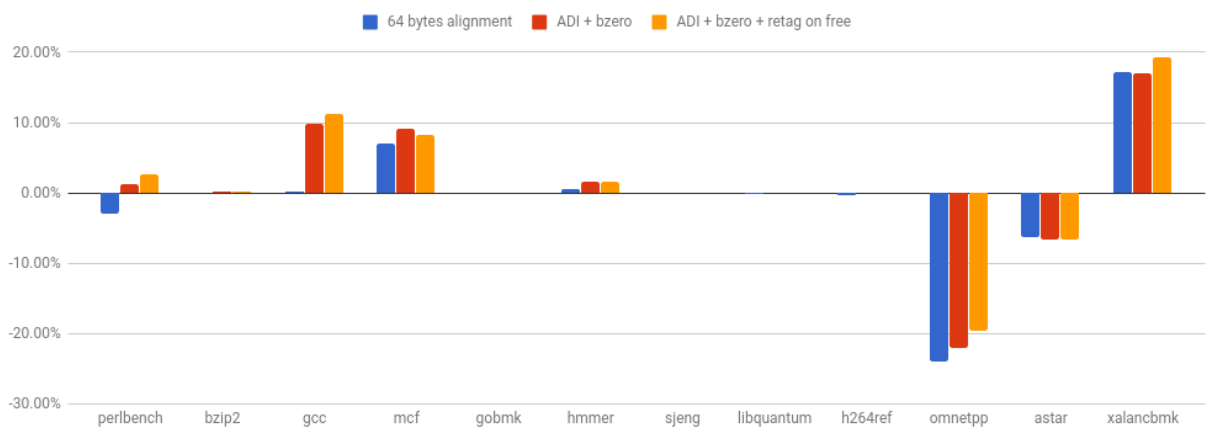

We have also measured the slowdown introduced by enabling the ADI's precise trap mode for stores and confirmed that this mode is not suitable for always-on usage in production.

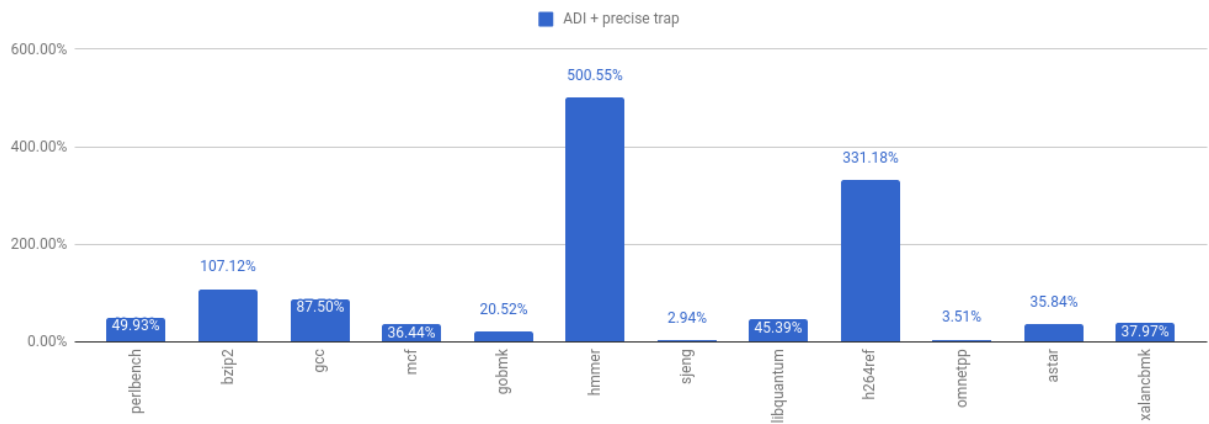

One more benchmark we used was the clang compiler: ADI slows it down by ~4% in imprecise mode and by 26% in precise.

**Conclusion:** a fully hardware assisted memory tagging (in heap-only mode) could have near-zero overhead except for very malloc-intensive applications, where the overhead will still remain moderate. Another important observation is that zeroing out the memory may have no additional cost if we are already tagging that memory.

## Code Size

The major source of **code size overhead** is the extra instructions to tag stack variables. SPARC toolchains do not support stack instrumentation. The current version of HWASAN has a naive stack instrumentation which increases the code size by 40-50%.
In HWASAN, tagging each stack variable requires several instructions:

```
eor    x19, x29, x29, lsr #20   << Get a semi-random tag from FP
mov    x8, sp                    << Get the address of the local variable
orr    x0, x8, x19, lsl #56     << Tag the address
lsr    x20, x8, #4               << Compute the shadow memory address
strb   w19, [x20]                << Tag the memory
```

A proper hardware implementation may require fewer, probably two or three, instructions, so we expect the code size increase to be more moderate. Still, compilers need to learn how to avoid excessive stack frame tagging by proving that certain address-taken local variables are immune to temporal and spatial bugs.

## Usage Modes

In this section we describe the main usage modes for memory tagging systems and share Google's prior experience with ASAN in similar circumstances.

## Testing

A system based on MT can be seen as a better ASAN (but: see "[Possible Improvements](#)"):
- Much smaller RAM overhead.
- More reliable detection of temporal bugs (on heap and stack). Unlike ASAN, MT does not rely on a quarantine to detect temporal bugs, and the probability of detecting such bugs does not drop over time. It is still possible to use small quarantines with MT to allow non-probabilistic detection of accesses to very recently freed memory.
- More reliable detection of some spatial bugs (non-linear buffer overflows or arbitrary memory accesses). Unlike ASAN, MT does not rely on redzones and so non-linear buffer overflows are detected with the same high probability.
- For fully hardware-assisted systems, i.e. for SPARC ADI but not for HWASAN:
  - Smaller CPU and Code Size overhead.
  - Finds buggy accesses in non-instrumented code or inside system calls, without requiring extra complexity in software.

While these improvements are very welcome, they may not provide enough motivation to implement full MT in hardware - a much simpler top-byte-ignore feature, if implemented in other CPU architectures, would provide 80% of benefits for *regular testing* at 20% of cost.

However, we believe that the other usage modes make it critical to have full MT in hardware.

## Always-on Bug Detection In Production

We have strong evidence that testing does not find *all* memory safety bugs that *actually happen in production*.

Despite our large-scale efforts for fuzzing the Chrome browser ([[1]](#), [[2]](#)) we keep finding memory safety bugs in the **shipped versions of Chrome**. A version of Chrome with [SyzyASAN instrumentation](#) used by a small subset of real users on Windows (so-called "canary channel") finds [two-three bugs every week](#) (not all of these bugs are public, the total number after 5 years of testing is ~950). The majority of these bugs are heap-use-after-free (SyzyASAN does not find stack-related bugs). These are the bugs that happen when real users use the Chrome browser normally, i.e. not necessarily when someone is trying to attack it. There is a skew towards bugs that require complicated user interaction, since other kinds of bugs are more often detected during testing & fuzzing. Shipping such instrumented builds to more users leads to more bugs discovered per month.

Our experience with Google server-side applications is similar. Several production services set up *ASAN canaries* (dedicated small clusters running an ASAN build on real traffic). Despite our massive testing and fuzzing efforts these canaries regularly find memory safety bugs that evaded testing.

Last but not least, the Android platform has [adopted](#) ASAN as the main memory sanitization tool, [replacing](#) the slower Valgrind. However, ASAN's overhead continues to be problematic. For one, Android typically runs dozens of concurrent processes ("Android apps" and services) on constrained devices. At that point, the combined memory overhead starts to negatively affect the performance often making devices unusable. Besides, ASAN requires specially-built libraries, which usually overflow the limited system storage of Android devices, necessitating a non-standard directory layout that complicates practical use outside of a development environment. Even with these constraints, ASAN has been very successful in finding platform and app issues.

The large overhead of ASAN makes these efforts extremely painful, costly, and often prohibitively complex. But a full hardware implementation of MT would solve these problems.

A system similar to SPARC ADI allows shipping a binary to production (web service, desktop, mobile, etc - as long as the hardware supports MT) that will find the majority of memory safety bugs actually triggered on real inputs. Given our evaluation of the overhead of SPARC ADI we believe that many applications can run with MT always on.

For the Android ecosystem a full hardware implementation of MT would also have the benefit of being applicable to code not found in the stock platform (eg: all additional native code across OEM devices). This has significant advantages from both stability and security points of view. For example, external developers routinely use ASAN to find and report security bugs in Android (many of which can be found in the [Pixel Security Bulletins](#)). Being able to trivially extend this kind of testing to code that ships on non-Pixel devices will greatly increase the number of issues that are detected (and fixed) across the ecosystem.

## Sampling In Production

Some applications will not tolerate even a single-digit % overhead in RAM or CPU. The great advantage of the hardware-assisted memory tagging is that it allows sampling-based bug detection *for heap* on multiple levels.
- Per-device or per-VM sampling: the same binary is shipped to all devices/VMs, but the feature is enabled only on a small random subset of devices/VMs at any moment. When MT is disabled on the device the overhead is exactly zero.
- Per-process sampling: for multi-process applications, such as the Chrome browser, checking could be enabled on a subset of processes. Processes with disabled MT will have no penalty.
- Per-allocation sampling: malloc may choose to tag a random subset of allocations to avoid the overhead caused by malloc over-alignment and setting the memory tags. In this case, the RAM overhead from storing the memory tags will remain, but the other overheads will decrease proportionally to the sampling rate.

When also instrumenting the stack frames, it will be harder to implement sampling without maintaining multiple builds. Even if the instructions that tag local variables are correct when MT is disabled, every process will still have the code size overhead. The possibility to implement sampling for stack will depend on how compact the tagging instructions are.

## Security Hardening

Every Chrome release contains fixes for memory safety bugs reported to Google by external security researchers. Many of these bugs are found by sophisticated fuzzing or code inspection; we don't know how many of them happen during normal usage of Chrome as opposed to only being triggered by clever attacks. In other words, no amount of testing, including testing in production, will eliminate all bugs.

Our hypothesis is that always-on memory tagging in production will serve as a significant obstacle for attackers as it will increase the cost of attacks, and more importantly, reduce their reliability and stealthiness.

Below we outline some strengths and weaknesses of the general MT scheme as an exploit mitigation. We invite security practitioners to study this subject in greater detail and suggest improvements to the general scheme.

### Strengths

- MT prevents the root cause of many classes of attacks (as opposed to other mitigations, e.g. CFI, that prevent the consequences of memory corruption).
- Attackers sensitive to having their exploits detected and reported will be forced to use only the subset of vulnerabilities/exploit techniques unaffected by MT to evade the mitigation.
- MT can provide bug reports to the vendor with enough information to allow timely fixes (unlike most other mitigations).
- Leaking one pointer's tag does not compromise any other tags (unlike with ASLR where leaking one pointer allows the attacker to determine the layout of an entire section of memory).
- The memory tags are hard or impossible to leak (depends on the actual hardware implementation and whether it's protected from side channel attacks similar to Meltdown).
- Provides full mitigation against uses of uninitialized memory, see below.

### Weaknesses

- Probabilistic nature of bug detection. Given a sufficient number of attempts MT can be bypassed (but if an exploit requires a chain of bugs, those low probabilities will multiply for every bug detected by MT).
- The address tag is stored in the upper pointer bits, potentially allowing the attacker to change the address tag via an integer overflow on the address. Protecting from this attack will require more compiler instrumentation.
- The address tag is stored in a fixed position in the high bits of the pointer. A buffer overflow not prevented by MT (e.g. an intra-object overflow) on little-endian processors could change the low bits of a pointer without changing the tag so that it continues to point into the same allocation (i.e. with the same tag) but increase the attacker's access in some way.
- The address tag may be leaked and/or modified by some remaining unprotected memory corruption classes, such as intra-object-buffer-overflow or type confusion. But if the attacker has those primitives, in many cases they won't need to bypass MT at all.
- Since the accesses to the memory and their tags are not atomic with respect to each other, racy use-after-frees may fail to be detected (but observing a racy use-after-free is probabilistic anyway).
- If the attacker can leak pointers (and hence their tags) for arbitrary objects they may be able to repeatedly cause the allocation/deallocation of new heap/stack objects until they've found two with matching tags to use in a use-after-free, linear overflow, etc.

## Legacy Code

Hardware memory tagging (e.g. SPARC ADI), allows testing legacy code without recompiling it -- only changes in malloc/free are required. This mode only allows finding heap-related bugs. Mixed modes are also possible where parts of the code are instrumented (e.g. to also find stack-related bugs) and other parts are not.

## Kernel

This document mostly discusses memory safety of user-space applications. However everything here equally applies to the OS kernels, such as Linux. We have been testing Linux with KASAN (Kernel-ASAN) for several years and the situation with memory safety there is at least as bad as in user space. KASAN has found at least 400 heap-use-after-free, 250 heap-buffer-overflow, and 5 stack-buffer-overflow bugs ([1], [2], [3], also: git log | grep KASAN). These bugs indicate that the Linux kernel will benefit from low-overhead memory tagging, optionally with per-machine or per-allocation sampling.

Besides, using the MT for user space applications will require OS kernel support. For example, HWASAN needs Linux to strip tags from user space pointers passed as system call arguments; the [patches](#) will be sent upstream shortly.

## Uninitialized Memory

Uses of uninitialized memory is another important, yet frequently overlooked, memory safety problem (e.g. [CVE-2017-1000380](#) and [CVE-2017-14954](#) in Linux, [more Linux info leaks](#), [Linux privilege escalations](#), [700+ bugs](#) in Chrome, [info leaks in Yahoo](#), [100+ bugs](#) found by oss-fuzz). Detecting uses of uninitialized memory is [often not trivial](#), but mitigating them is. The compiler and run-time should simply initialize all memory on allocation. Typically, this is considered to be not worth the overhead, but if we already tag the memory, initializing the same memory may come for free.

Our evaluation of SPARC ADI demonstrates that initializing the heap memory to zero while tagging it has no additional cost. This important property allows applications that use always-on MT to also mitigate uses of uninitialized heap memory (and for stack memory too, if the stack is instrumented).

# Possible Improvements

## Precision Of Buffer Overflow Detection

A disadvantage of the memory tagging scheme described here compared to ASAN and similar tools is the coarse precision of buffer overflow detection. Example:

```
// Assuming TG=16
uint8_t *p = (uint8_t*)malloc(10);
p[12] = 0; // same 16-byte granule, goes undetected
p[16] = 0; // will be detected
```

The bigger the TG, the bigger this disadvantage is. Some software strategies may soften the problem to some extent (e.g. periodically allocating objects right-aligned within the granule, instead of left-aligned, where permitted by the ABI) but won't eliminate it. Many such bugs will remain undetected.

One way to support better buffer overflow precision is to use more bits in the memory tag to represent the size within the granule: $\log_2(TG)$ bits are needed to represent the full size, fewer bits could provide a compromise in detection precision. However, this will use too many precious memory tag bits.

A more complicated scheme may require to reserve one memory tag value to indicate that the first N bytes (0<N<TG) of the granule are valid and the rest TG-N bytes are not. Additional information (e.g. N and the real memory tag to match against the pointer tag) could be stored in the right-most byte(s) of the granule. It's unclear how viable this idea is for full hardware implementations.

For applications that always use MT this is less of a problem because such *intra-granule* overflows will not cause memory corruption or information leaks (but will remain logical bugs).

### Probability Of Bug Detection

The probability of bug detection with MT primarily depends on the tag size. With TS=4, the probability is borderline tolerable (15/16=94%), with TS=8 it is already very good (255/256 = 99.6%). But larger values of TS incur larger RAM overheads. Clever software strategies will need to be invented to increase the probability of bug detection with small values of TS.

## Conclusion

This paper describes memory tagging (MT) - a general scheme for hardware-assisted C/C++ memory safety, and evaluates two implementations: SPARC ADI (full hardware assistance) and AArch64 HWASAN (minimal hardware assistance).

We discuss the possible usage modes for memory tagging and potential improvements. The major benefit from MT is finding bugs in production, either with always-on MT or by sampling. Traditional pre-production testing and fuzzing will also benefit from less costly and more reliable bug detection.

Memory tagging will not eliminate all memory safety bugs; however, our analysis indicates that memory tagging, when widely supported by hardware, will help significantly reduce the number of such bugs and is likely to complicate exploitation of the few remaining ones.

We call for a wider discussion of memory tagging and encourage the CPU and OS vendors to support it.